\documentclass[twocolumn,floatfix,superscriptaddress,a4paper,showpacs,showkeys,nofootinbib,reprint,prc]{revtex4-1}

\usepackage{epsfig}
\usepackage{latexsym}
\usepackage{xspace}
\usepackage[colorlinks=true,linktocpage=true,linkcolor=blue,citecolor=blue,allcolors=blue]{hyperref}
\usepackage[utf8]{inputenc}
\usepackage{indentfirst}
\usepackage{enumerate}
\usepackage{color}
\usepackage{tabularx}

\usepackage{amsmath}
\usepackage{amssymb}
\usepackage[english]{babel}
\usepackage{url}
\newcommand{\eq}[1]{\begin{align} #1 \end{align}}

\begin{document}


\title{
Finite resonance widths influence the thermal-model description of hadron yields 
}

\author{Volodymyr Vovchenko}
\affiliation{
Institut f\"ur Theoretische Physik,
Goethe Universit\"at Frankfurt, Max-von-Laue-Str. 1, D-60438 Frankfurt am Main, Germany}
\affiliation{Frankfurt Institute for Advanced Studies, Giersch Science Center, Goethe Universit\"at Frankfurt, Ruth-Moufang-Str. 1, D-60438 Frankfurt am Main, Germany}

\author{Mark I. Gorenstein}
\affiliation{Frankfurt Institute for Advanced Studies, Giersch Science Center, Goethe Universit\"at Frankfurt, Ruth-Moufang-Str. 1, D-60438 Frankfurt am Main, Germany}
\affiliation{Bogolyubov Institute for Theoretical Physics, 03680 Kiev, Ukraine}

\author{Horst Stoecker}
\affiliation{
Institut f\"ur Theoretische Physik,
Goethe Universit\"at Frankfurt, Max-von-Laue-Str. 1, D-60438 Frankfurt am Main, Germany}
\affiliation{Frankfurt Institute for Advanced Studies, Giersch Science Center, Goethe Universit\"at Frankfurt, Ruth-Moufang-Str. 1, D-60438 Frankfurt am Main, Germany}
\affiliation{
GSI Helmholtzzentrum f\"ur Schwerionenforschung GmbH, Planckstr. 1, D-64291 Darmstadt, Germany}

\begin{abstract}
Different scenarios for modeling resonances in a thermal model description of hadron yields measured in heavy-ion collisions are explored: the zero-width approximation, the energy independent Breit-Wigner scheme, and the energy dependent Breit-Wigner~(eBW) scheme.
Application of the eBW scheme leads to a notable suppression in the proton yields, stemming mainly from a reduced feeddown from $\Delta$ resonances because of the threshold effects.
A significantly improved agreement of thermal model with hadron yields measured in Pb-Pb collisions at $\sqrt{s_{_{NN}}} = 2.76$~TeV by the ALICE collaboration is obtained in the eBW scheme at $T \simeq 155$~MeV, indicating a possible resolution of the so-called 'proton anomaly'.
The results obtained show that there are significant systematic uncertainties in the thermal model due to the modeling of broad resonances.
\end{abstract}

\pacs{24.10.Pa, 25.75.Gz}

\keywords{thermal model, particle production, resonance widths}

\maketitle


\section{Introduction}

Thermal
models have long been used to estimate properties of the medium created in relativistic heavy-ion collisions~\cite{Mekjian:1977ei,Gosset:1988na,Mekjian:1978us,Stoecker:1981za,Csernai:1986qf,Hahn:1987tz,Hahn:1986mb,Cleymans:1992zc,BraunMunzinger:1996mq,Becattini:2000jw}.
These  models  assume that the emitted particles stem from a statistically equilibrated system,
conventionally described as an ideal hadron resonance gas (HRG) of all known hadrons and resonances.
The temperature $T$ and baryon chemical
potential $\mu_B$ are obtained by fitting the observed yields of hadrons, and a surprisingly good description is achieved across a broad range of collision energies~(see, e.g., \cite{Braun-Munzinger:2015hba} for an overview).

A considerable attention have been devoted to the hadron yield data measured by the ALICE collaboration in Pb-Pb collisions at $\sqrt{s_{_{NN}}} = 2.76$~TeV.
There, already the usual ideal gas implementation of the thermal model works relatively well~\cite{Stachel:2013zma,Petran:2013lja,Floris:2014pta}, even for the loosely bound states 
such as the light nuclei.
One notable exception are the yields of protons and antiprotons, which are overestimated substantially
by thermal model relative to other hadron yields. This has been dubbed as the 'proton anomaly'~\cite{Andronic:2012dm,Stachel:2013zma}, and various mechanisms, such the annihilation in the hadronic phase~\cite{Steinheimer:2012rd,{Becattini:2012xb,Becattini:2016xct}},
incomplete hadron spectrum~\cite{Stachel:2013zma,Noronha-Hostler:2014aia},
chemical non-equilibrium~\cite{Petran:2013lja,{Begun:2013nga,Begun:2014rsa}}, flavor dependency in the freeze-out temperature~\cite{Bellwied:2013cta,Chatterjee:2016cog} or in hadron eigenvolumes~\cite{Alba:2016hwx} have been suggested to explain it.

In the present work we investigate the little explored but important aspect of thermal models -- the treatment of the finite resonance widths.
In the literature these are most commonly treated either in the zero-width approximation, or by applying the additional integration over the Breit-Wigner shape of a resonance with an energy independent width parameter.
The latter prescription is used, e.g., in the Florence code~\cite{Becattini:1995if} or in the \texttt{THERMUS} package~\cite{Wheaton:2004qb}.
Energy-dependent Breit-Wigner shapes are considered in the~\texttt{SHARE} package~\cite{Torrieri:2004zz}.
To our knowledge, no comparative study between these different possibilities has been considered so far for hadron yields at the LHC.
Here we demonstrate that modeling finite resonance widths has an important effect on final hadron yields, in particular on the proton yield.

The manuscript is organized as follows.
Different scenarios for modeling finite resonance widths are described in Sec.~\ref{sec:models}.
Their influence on final hadron yields and on thermal fits to the ALICE data are discussed, respectively, in Sec.~\ref{sec:yields} and \ref{sec:fits}.
Concluding remarks in Sec.~\ref{sec:summary} close the article.

\section{Model description}
\label{sec:models}

\subsection{HRG model in zero-width approximation}

In its simplest version, the thermal model assumes thermally and chemically equilibrated non-interacting gas of known hadrons and resonances at chemical freeze-out.
The primordial hadron densities at the chemical freeze-out are then given by the corresponding Fermi-Dirac or Bose-Einstein distribution functions.
Resonances are included as non-interacting point-like particles with zero width.
The primordial hadron yields read
\eq{\label{eq:Nprim}
\langle N_i^{\rm prim} \rangle & = \frac{d_i V}{2\pi^2} \, \int_0^{\infty} \frac{k^2 dk} {\exp\left(\frac{\sqrt{k^2+m_i^2} - \mu_i}{T}\right)+\eta_i}~.
}
Here $V$, $T$, and $\mu_i$
are the system volume, temperature,
and chemical potential of $i$th particle species
at freeze-out, 
$m_i$, and $d_i$ are, respectively, 
mass and degeneracy factor of $i$th particle, 
$\eta_i = +1$ for fermions and $\eta_i = -1$ for bosons.

The final mean multiplicity $\langle N_i^{\rm tot}\rangle$ of $i$th
particle species  is calculated in thermal models
as the sum of the primordial mean multiplicity
$\langle N^{*}_i\rangle \equiv n_i \, V$ and of resonance decay
contributions as follows
\eq{\label{eq:Ntot}
\langle N_i^{\rm tot} \rangle~ =~
\langle N^{\rm prim}_i\rangle~ +~ \sum_R \langle n_i \rangle_R \, \langle N^{\rm prim}_R\rangle~,
}
where $\langle n_i \rangle_R$ is the average number of particles
of type $i$ resulting from decay of resonance $R$. $\langle n_i \rangle_R$ includes contribution from both, the direct decays of resonance $R$ resulting in the production of hadron $i$, as well as the contributions resulting from the chain of decays via lower-mass resonances.

\subsection{Energy independent Breit-Wigner}

A simple way to model the finite resonance widths is to assume  the
Breit-Wigner (BW) distribution around the pole mass with an energy independent width parameter~(see, e.g., Refs.~\cite{Becattini:1995if,Wheaton:2004qb}).
This corresponds to the following modification of Eq.~\eqref{eq:Nprim}:
\eq{\label{eq:NprimBW}
\langle N_i^{\rm prim} \rangle & = \frac{d_i V}{2\pi^2} \, \int_{m_i^{\rm min}}^{m_i^{\rm max}} d m \, \rho_i^{\rm BW}(m) \int_0^{\infty} k^2 dk \nonumber \\
& \times
\left[ \exp\left(\frac{\sqrt{k^2+m^2} - \mu_i}{T}\right)+\eta_i\right]^{-1}~.
}
Here $\rho_i^{\rm BW}(m)$ is the relativistic Breit-Wigner distribution\footnote{Application of the non-relativistic Breit-Wigner distribution instead of the relativistic one leads to very similar results.}
\eq{\label{eq:rhoBW}
\rho_i^{\rm BW}(m) = A_i \, \frac{2 \, m \, m_i \, \Gamma_i}{(m^2-m_i^2)^2 + m_i^2 \Gamma_i^2},
}
where $\Gamma_i$ is the width of the resonance $i$, and $A_i$ is the normalization constant fixed by the condition $\int_{m_i^{\rm min}}^{m_i^{\rm max}} d m \, \rho_i^{\rm BW}(m) = 1$.
To fix the integration limits, $m_i^{\rm min}$ and $m_i^{\rm max}$, we conservatively assume that the mass of the resonance is distributed in the $\pm 2 \Gamma_i$ interval around the pole~\cite{Becattini:1995if,Wheaton:2004qb}, but also take into account the decay threshold mass $m_i^{\rm thr}$.
Therefore, $m_i^{\rm min} = \operatorname{max}(m_i^{\rm thr}, m_i - 2\,\Gamma_i)$ and $m_i^{\rm max} = m_i + 2\,\Gamma_i$.
The threshold mass $m_i^{\rm thr}$ is computed as a weighted average of individual threshold masses over all decay channels of resonance $i$, the branching ratios being the corresponding weights.

The final hadron yields in the energy independent BW scheme are computed according to Eq.~\eqref{eq:Ntot}.
Note that for $\Gamma_i\rightarrow 0$ it follows $\rho_i^{\rm BW}\cong \delta(m-m_i)$, and Eq.~(\ref{eq:NprimBW}) is reduced to Eq.~(\ref{eq:Nprim}).

\subsection{Energy dependent Breit-Wigner}

The simple energy independent BW scheme does not capture correctly the threshold effects.
These may be 
quite significant for broad resonances.
The low-mass tail of the resonance may give a 
significant contribution to thermodynamics due the Boltzmann factor.
A better theoretically motivated framework to treat resonances is the energy dependent Breit-Wigner~(eBW) scheme, where the partial widths $\Gamma_{i \to j} (m)$ of different decays of the resonance $i$ are mass dependent.

There are different prescriptions available in the literature to compute $\Gamma_{i \to j} (m)$~(see, e.g., \cite{Manley:1992yb}).
In this work we apply the simple, two-body decay motivated prescription, which is similar to the one used in the \texttt{SHARE}~\cite{Torrieri:2004zz} package:
\eq{\label{eq:GammaijeBW}
\Gamma_{i \to j} (m) = b_{i \to j}^{\rm pdg} \, \Gamma_i^{\rm pdg} \, 
\frac{\left[ 1 - \left(\frac{m_{i \to j}^{\rm thr}}{m}\right)^2 \right]^{L_{i \to j} + 1/2}}{\left[ 1 - \left(\frac{m_{i \to j}^{\rm thr}}{m_i}\right)^2 \right]^{L_{i \to j} + 1/2}}
}
valid for $m > m_{i \to j}^{\rm thr}$~(and equal to zero otherwise).
Here $m_{i \to j}^{\rm thr}$ is the sum of the masses of all decay products in the channel $j$. If one of the daughter particles is a resonance itself, then its pole mass is taken for calculating $m_{i \to j}^{\rm thr}$, for simplicity.
$b_{i \to j}^{\rm pdg}$ and $\Gamma_i^{\rm pdg}$ are, respectively, the branching ratio for channel $j$ and the total width, as listed in the Particle Data Tables.
$L_{i \to j}$ is the angular momentum released in the decay channel $j$.
Following the implementation in the \texttt{SHARE} package, this quantity is calculated here as $L_{i \to j} = |J_{i} - \sum_{k}{J_k}|$, where $J_i$ is the total angular momentum of particle species $i$ and the index $k$ runs over all daughter particles in the decay channel $j$.

The denominator in Eq.~\eqref{eq:GammaijeBW} is the normalization factor, which ensures that, at the pole mass $m = m_i$, the sum of all partial widths is equal to the total PDG width, $\Gamma_i^{\rm pdg}$, and that
the ratios of the partial widths to the total width are equal to the PDG branching ratios, $b_{i \to j}^{\rm pdg}$.

The total energy dependent width in the eBW scheme reads
\eq{\label{eq:GammaeBW}
\Gamma_i (m) = \sum_j \Gamma_{i \to j} (m),
}
and the mass distribution is
\eq{\label{eq:rhoeBW}
\rho_i^{\rm eBW}(m) = A_i \, \frac{2 \, m \, m_i \, \Gamma_i(m)}{(m^2-m_i^2)^2 + m_i^2 [\Gamma_i(m)]^2}.
}

The decay branching ratios in the eBW scheme are given by the ratios of the corresponding partial widths to the total widths.
They are 
therefore 
energy dependent:
\eq{
b_{i \to j} (m) = \frac{\Gamma_{i \to j}(m)}{\Gamma_i (m)}.
}
This energy dependence is taken into account here for decays of the primordial resonances.
The implication is that the average number $\langle n_i \rangle_R$ of particles of type $i$ resulting from a decay of resonance $R$ is 
an energy dependent quantity itself, i.e. $\langle n_i \rangle_R \to \langle n_i \rangle_R (m)$.
This results in the following modification of Eq.~\eqref{eq:Ntot} for calculating the final hadron yields in the eBW scheme:
\eq{\label{eq:NtoteBW}
\langle N_i^{\rm tot, eBW} \rangle & =
\int d m \frac{d \langle N^{\rm prim}_i\rangle ~ (m)}{dm}
\nonumber \\
& \quad +
\sum_R \int d m \, \langle n_i \rangle_R (m) \, \frac{d \langle N^{\rm prim}_R \rangle ~ (m)}{dm}~,
}
with
\eq{
\frac{d \langle N^{\rm prim}_i\rangle ~ (m)}{dm} = \rho_i^{\rm eBW}(m) \, \frac{d_i V}{2\pi^2} \, \int_0^{\infty} \frac{k^2 dk} {\exp\left(\frac{\sqrt{k^2+m^2} - \mu_i}{T}\right)+\eta_i}~.
}

Our calculations suggest that the modification $\langle n_i \rangle_R \to \langle n_i \rangle_R (m)$ leads to very small differences in the final observables. 
To a good approximation, the branching ratios $b_{i \to j}$ can be considered energy independent even in the eBW scheme.

\subsection{Some other details}

The hadron list employed in the present work conservatively includes only the established hadrons and resonances listed in the 2014 edition of the Particle Data Tables~\cite{Agashe:2014kda}, as well as light nuclei up to $^4 {\rm He}$. We do not consider here charm and bottom flavored hadrons.
The decays and the corresponding branching ratios are also taken from PDG.
For some established resonances the decay channels are not well known.
In this case we assign the missing decays based on a similarity to the known decay channels of similar resonances with the same quantum numbers.
This procedure is a minor new element compared to our earlier studies~\cite{Vovchenko:2015cbk,Vovchenko:2015idt}.

The feeddown contributions include strong and electromagnetic decays, which matches the experimental conditions of the ALICE experiment at the LHC.

The finite resonance widths modeling is not applied for very narrow resonances for which $\Gamma_i / m_i < 0.01$.
Therefore, for all these narrow resonances (and also for all stable hadrons)
\eq{
\rho_i^{\rm BW} = \rho_i^{\rm eBW} = \delta(m - m_i), \qquad i \in \text{stable/narrow}
}

The calculations are performed within the open source \texttt{Thermal-FIST} package~\cite{ThermalFIST}, where the eBW scheme is implemented.

\section{Influence on hadron yields}
\label{sec:yields}

\begin{figure}[t]
  \centering
  \includegraphics[width=.49\textwidth]{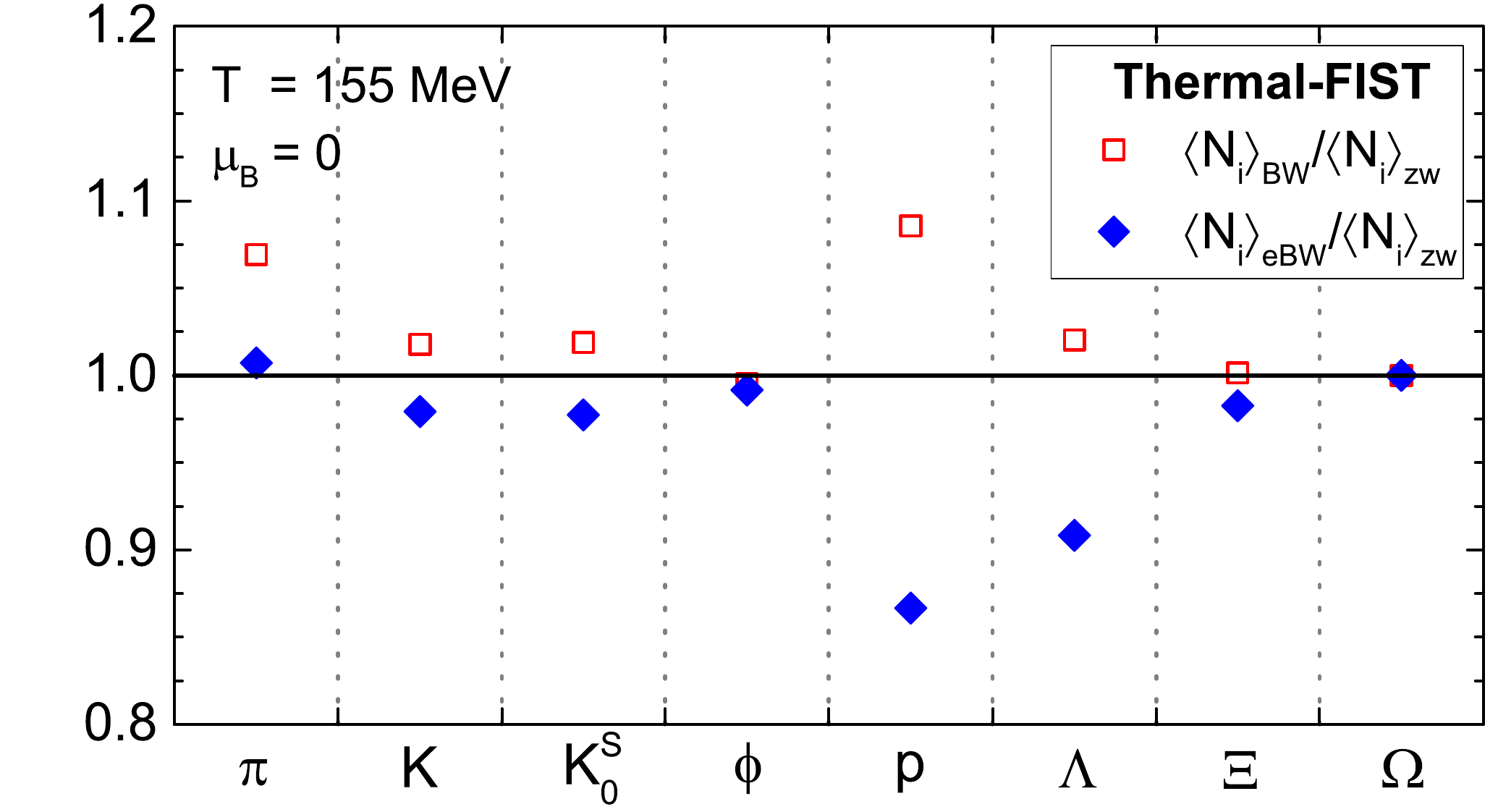}
  \caption{The ratios of the final hadron yields calculated within the BW~(open red squares) and eBW schemes~(solid blue diamonds) over the ones calculated in the zero-width approximation at $T = 155$~MeV and $\mu_B = 0$.
  }
  \label{fig:ratios}
\end{figure}

To illustrate how different schemes for modeling resonances affect the experimentally observable final hadron yields we consider the ratios of the final hadron yields in the BW and eBW schemes to the ones in the zero width approximation.
The corresponding quantities are shown in Fig.~\ref{fig:ratios}, calculated for $\pi^{\pm}$, $K^{\pm}$, $K_0^S$, $\phi$, (anti)protons, (anti)$\Lambda$, (anti)$\Xi^-$, and (anti)$\Omega$ at $T = 155$~MeV and $\mu_B = 0$.
These $T$ and $\mu_B$ values correspond to the chemical freeze-out conditions in Pb-Pb collisions at LHC obtained from the thermal fits in the ideal HRG model.
Note that at $\mu_B = 0$  it follows $\mu_i = 0$ 
for all hadron species, one therefore obtains identical yields for particles and antiparticles.

The effect of finite resonance widths in the BW scheme is to enhance the yields of hadrons relative to the zero width approximation.
The reason is the integration of the BW distribution with the Boltzmann factor, which normally yields an effective average mass of a resonance which is smaller than its pole mass.
This leads to an enhanced feeddown to the final yields of hadrons listed above.
This affects more strongly the feeddown from broad resonances such as $\Delta$'s and $N^*$'s, that
is why the strongest enhancement in final yields is seen in Fig.~\ref{fig:ratios} for pions and protons.

\begin{figure*}[!ht]
  \centering
  \includegraphics[width=.49\textwidth]{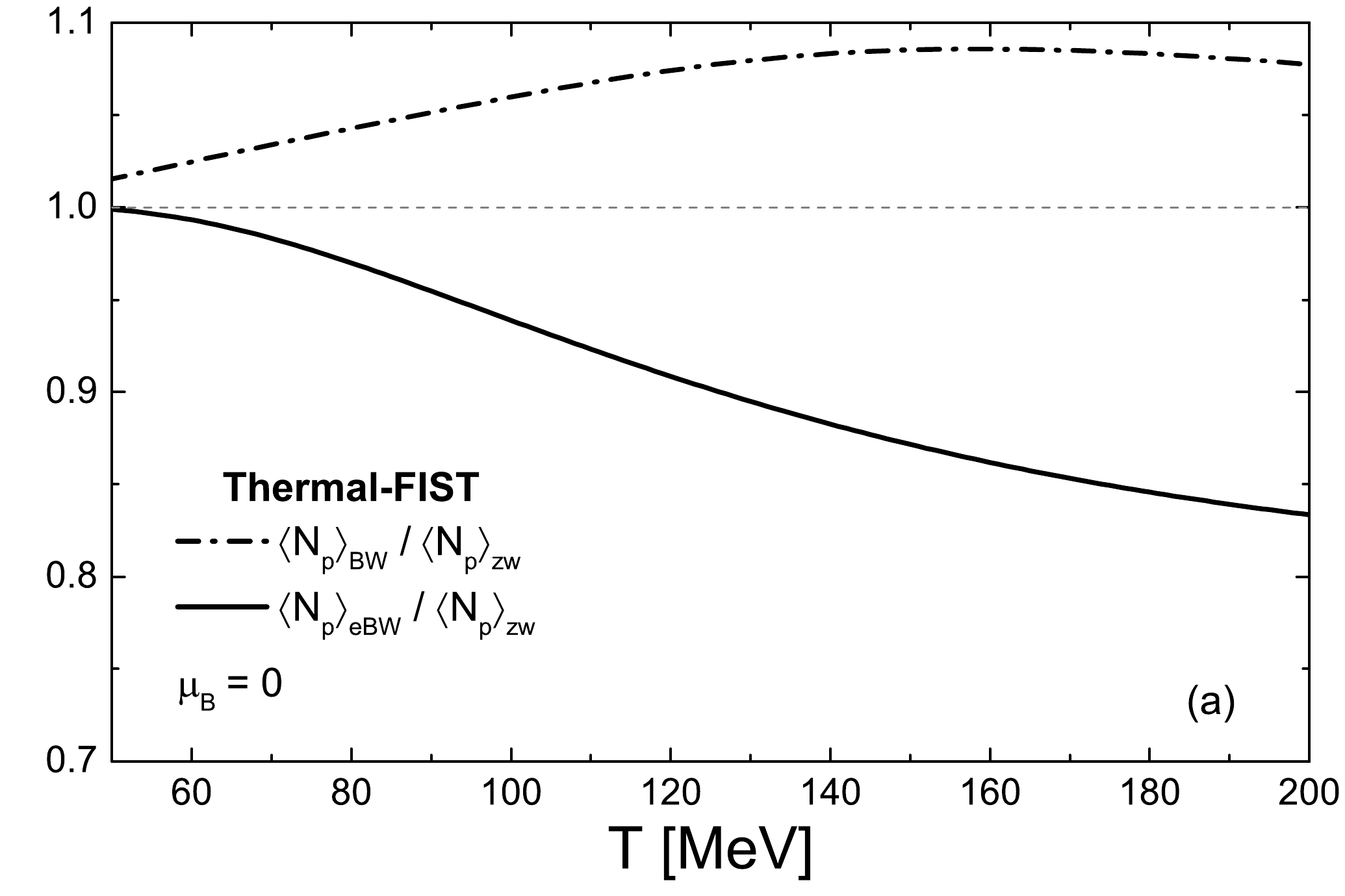}
  \includegraphics[width=.49\textwidth]{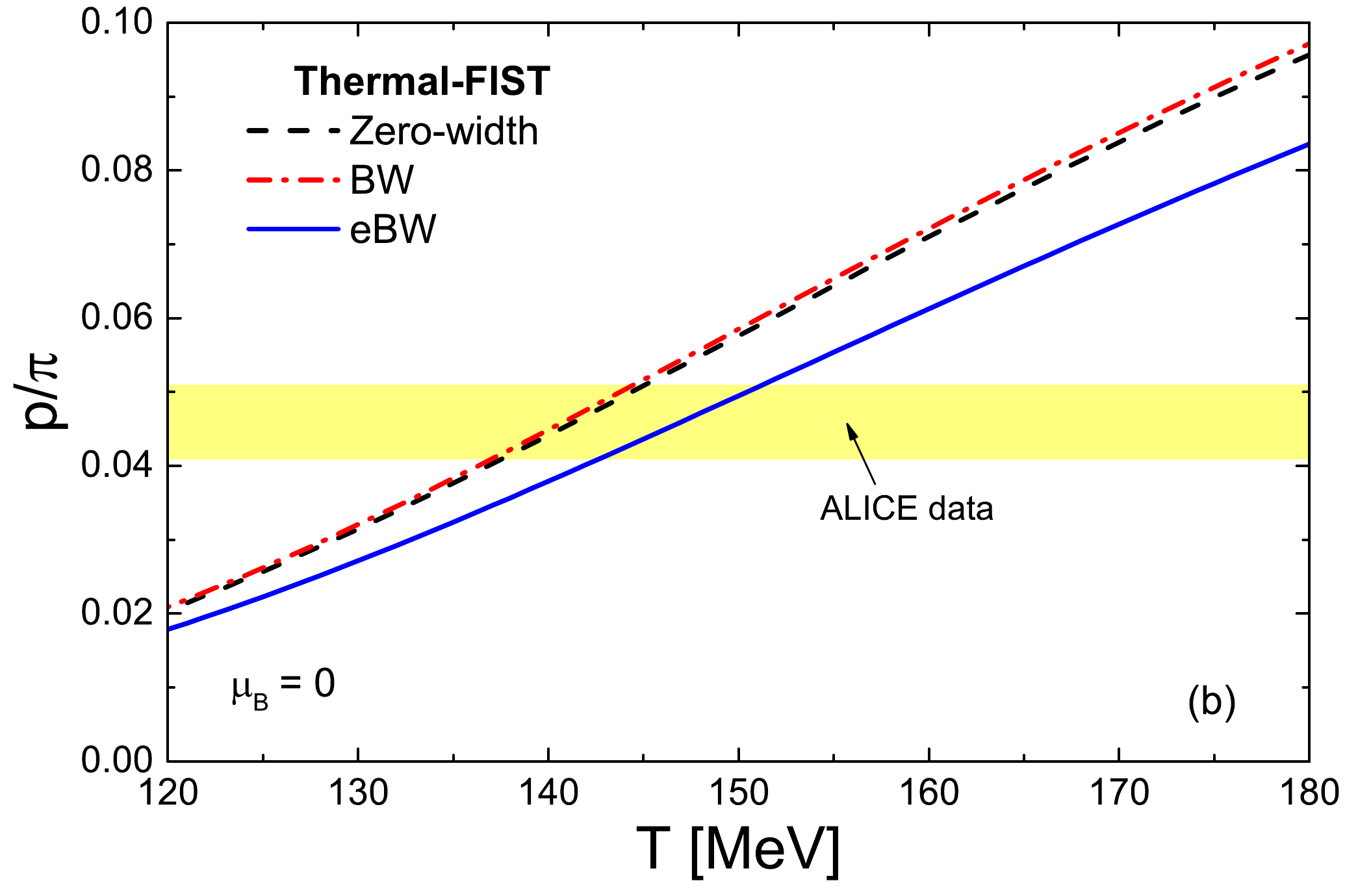}
  \caption{(a) Temperature dependence of ratio of the final proton yield calculated within the BW~(dash-dotted line) and eBW~(solid line) over the final proton yield in the zero width approximation.
  (b) Temperature dependence of the proton-to-pion ratio calculated within the HRG model in zero width approximation~(dashed line), BW scheme~(dash-dotted) line, and the eBW scheme~(solid line). The yellow band depicts the proton-to-pion ratio measured by the ALICE collaboration in 0-10\% central Pb-Pb collisions at $\sqrt{s_{_{NN}}} = 2.76$~TeV.
  }
  \label{fig:protons}
\end{figure*}

The eBW scheme leads to a different effect -- a suppression instead of enhancement --
a mild one for mesons, $\Xi$'s and $\Omega$'s, and a stronger one for $\Lambda$'s and protons.
The largest  effect is seen in the suppression of the final proton yield, where the reduction factor is equal to about 0.86.

To understand this effect one can consider feeddown contributions to the final proton yield.
In the zero width case, the fraction of final protons coming from decays of various $\Delta$ resonances is about 40\%.
In the eBW scheme, this feeddown is reduced by about 25\%.
The reason for that are significant threshold effects for $\Delta$'s, owing to their large widths.
These threshold effects suppress yields of $\Delta$'s relative to the zero width case quite notably.

Figure~\ref{fig:protons}(a) shows the temperature dependence of the modification of the proton yield relative to the zero-width approximation, calculated at $\mu_B = 0$ for the BW and eBW schemes.
The deviations from unity are smaller for lower temperatures, where the dominant contributions to the final proton yields come from primordial protons.
The effects of finite widths, however, are notable already at temperatures as low as $T \simeq 100$~MeV.

Figure~\ref{fig:protons}(b) depicts the temperature dependence of the $p/\pi$ ratio in the three considered schemes.
The experimental value $\langle p \rangle / \langle \pi \rangle \approx 0.046 \pm 0.005$, measured by the ALICE collaboration in the 0-10\% most central Pb-Pb collisions at $\sqrt{s_{_{NN}}} = 2.76$~TeV~\cite{Abelev:2013vea}, is shown by the yellow band.
The $p/\pi$ ratio is very similar in the zero-width and BW schemes.
It crosses the experimental value at a rather low temperature value of $T = 141 \pm 4$~MeV.
This value is substantially lower than the freeze-out temperature of $T \simeq 155$~MeV, obtained from fits to all measured hadron yields~\cite{Floris:2014pta}, and is the main reason for the tension between thermal model and data.
In the eBW scheme, on the other hand, the $p/\pi$-vs-$T$  curve is notably lower than the other two, the experimental value is crossed at  $T = 147 \pm 4$~MeV, i.e. about 6~MeV higher.
One can therefore expect that some tension with the data is removed when energy dependent resonance widths are considered.
We will illustrate this in the next section by performing full thermal fits to ALICE data within the different schemes considered.

\section{Data analysis}
\label{sec:fits}

A significant effect of modeling finite resonance widths on final hadron yields implies that thermal fits are sensitive to this modeling.
Here we study this sensitivity by performing the thermal fits to the experimental data within different schemes for resonance widths.

\subsection{Experimental data set}

We analyze the midrapidity hadron yields measured by ALICE collaboration in Pb-Pb collisions at $\sqrt{s_{_{NN}}} = 2.76$~TeV~\cite{Abelev:2013vea,Abelev:2013xaa,Abelev:2014uua,ABELEV:2013zaa,Adam:2015vda,Adam:2015yta,Acharya:2017bso}.
Since the asymmetry in the mid-rapidity densities between particle and antiparticles is negligible at LHC, we use the symmetrized yields of particles and antiparticles.
Data at different centralities are considered, with a focus on the 0-10\% most central collisions. These yields are corrected for weak decays.

The experimental data for 0-10\% centrality are listed in Table~\ref{tab:data}.
The data includes yields of light nuclei, for completeness.
The $^3\textrm{He}$ yields for 0-10\% were obtained by rescaling the 0-20\% yields~\cite{Adam:2015vda} by using the $d$ based scaling factor 1.127~\cite{PBMCPOD2016}.
The $^3_\Lambda\textrm{H}$ yields are obtained assuming the 25\% branching ratio of the $^3_\Lambda\textrm{H} \to ^3 \textrm{He} + \pi$ decay~\cite{Adam:2015yta}.
In the present work we do not discuss whether the point-particle approximation is appropriate for light nuclei, and whether
the light nuclei should be considered in the thermal model at all, as an argument can be made that they should be formed via the coalescence mechanism instead. 
 
The data at different centralities includes the same hadrons as listed in Table~\ref{tab:data}, with the exception of light nuclei for which the data at different centralities are scarce.
The following centralities are considered:
0-10\%, 10-20\%, 20-40\%, 40-60\%, 60-80\%.

\begin{table}[t]
 \caption{The hadron midrapidity yields for $0-10$\% most central Pb+Pb collisions at $\sqrt{s_{_{NN}}} = 2.76$~TeV measured by the ALICE collaboration and used in the thermal fits in the present work.} 
 \centering                                                 
 \begin{tabular}{c|c|c}                                   
 \hline
 \hline
 Particle & Measurement ($dN/dy$) & Reference \\
 \hline
 $(\pi^+ + \pi^-)/2$       & $668.75 \pm 47.5$ & \cite{Abelev:2013vea} \\
 $(K^+ + K^-)/2$         & $99.75 \pm 8.25$  & \cite{Abelev:2013vea} \\
 $K^0_S$       & $100  \pm 8$  & \cite{Abelev:2013xaa} \\
 $\phi$       & $12.75  \pm 1.59$  & \cite{Abelev:2014uua} \\
 $(p+\bar{p})/2$           & $30.75  \pm 2.50$  & \cite{Abelev:2013vea} \\
 $\Lambda$     & $24.0  \pm 2.5$  & \cite{Abelev:2013xaa} \\
 $(\Xi^- + \bar{\Xi}^-)/2$       & $3.335  \pm 0.238$  & \cite{ABELEV:2013zaa} \\
 $(\Omega + \bar{\Omega})/2$       & $0.595 \pm 0.100$  & \cite{ABELEV:2013zaa} \\
\hline
 $(\textrm{d} + \overline{\textrm{d}})/2$       & $0.097 \pm 0.018$  & \cite{Adam:2015vda} \\
 $(^3\textrm{He} + ^3\overline{\textrm{He}})/2$       & $(2.85 \pm 0.76) \cdot 10^{-4}$  & \cite{Adam:2015vda} \\
 $(^3_\Lambda\textrm{H} + ^3_{\bar{\Lambda}}\overline{\textrm{H}})/2$       & $(1.47 \pm 0.42) \cdot 10^{-4}$  & \cite{Adam:2015yta} \\
 $(^4\textrm{He} + ^4\overline{\textrm{He}})/2$       & $(9.5 \pm 4.8) \cdot 10^{-7}$  & \cite{Acharya:2017bso} \\
\hline
\hline
 \end{tabular}
\label{tab:data}
\end{table}

\subsection{Most central collisions}

\begin{figure*}[t]
  \centering
  \includegraphics[width=.80\textwidth]{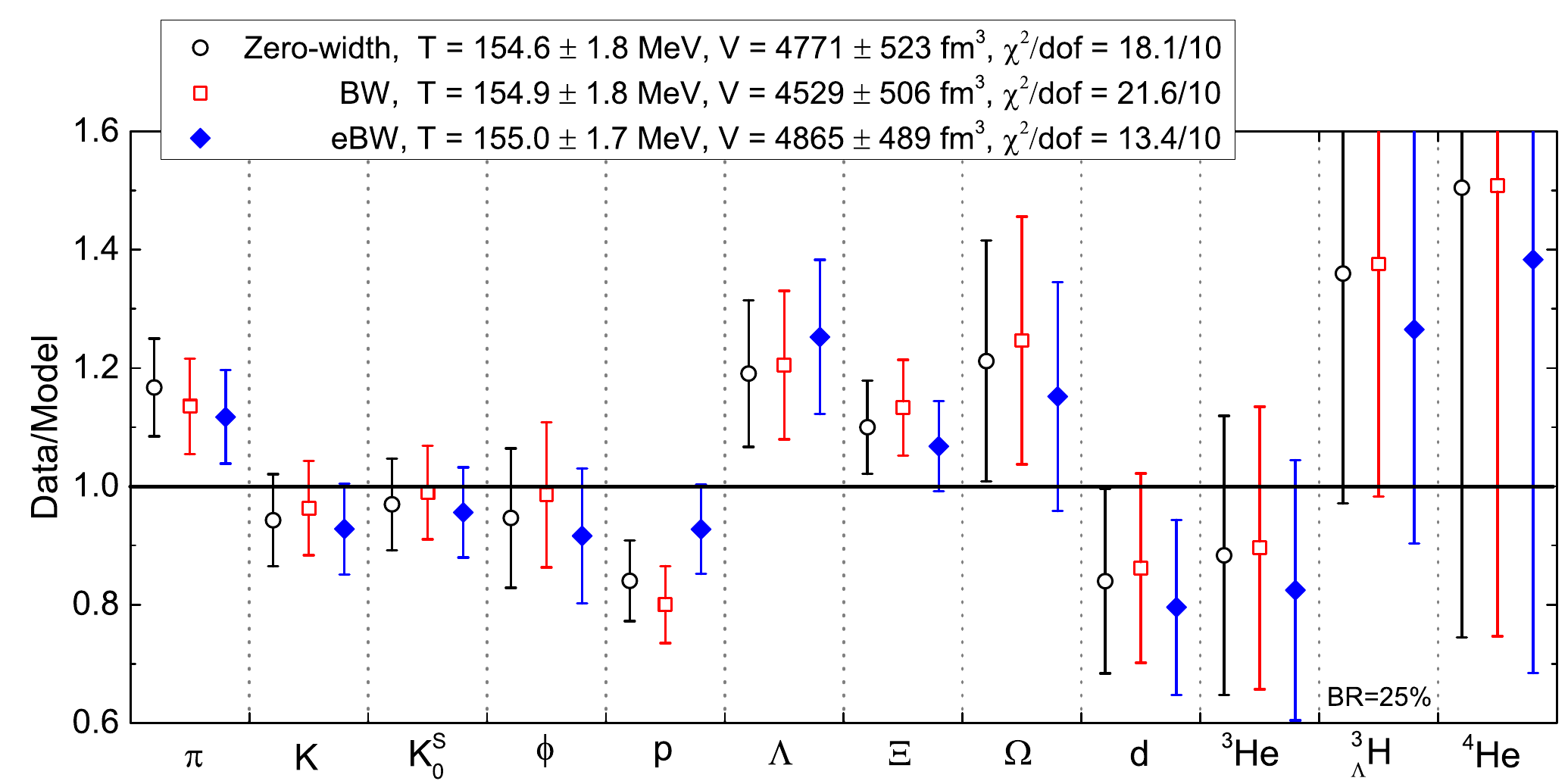}
  \caption{The data/model ratios resulting from thermal fits to particle yields measured in 0-10\% central Pb-Pb collisions at $\sqrt{s_{_{NN}}} = 2.76$~TeV, performed within the zero-width approximation~(open black circles), the BW scheme~(open red squares), and the eBW scheme~(full blue diamonds). The particle yields which were fitted are listed in Table~\ref{tab:data}.}
  \label{fig:fitscentral}
\end{figure*}

First, the hadron yield data for most central (0-10\%) Pb-Pb collisions at $\sqrt{s_{_{NN}}} = 2.76$~TeV, listed in Table~\ref{tab:data}, are analyzed.
The fits are performed within the \texttt{Thermal-FIST} package for ideal HRG model with the three scenarios for modeling finite resonance widths described in Sec.~\ref{sec:models}.
The yields symmetrized between particles and antiparticles are fitted, therefore $\mu_B = 0$ is enforced in the thermal model used.
The results of the fits are shown in Fig.~\ref{fig:fitscentral}, as the data/model ratios for all the fitted yields.

All three fits yield very similar values of the chemical freeze-out temperature $T \simeq 155 \pm 2$~MeV and the volume parameter $V \simeq 4700 \pm 500$~fm$^3$.
However, there are differences in the resulting $\chi^2 / {\rm dof}$ values.
The fit in the zero width approximation gives $\chi^2 / {\rm dof} = 18.1 / 10$ which is slightly better than the $\chi^2 / {\rm dof} = 21.6 / 10$ value obtained by fitting within the BW scheme.
The application of the eBW scheme leads to notably reduced value of $\chi^2 / {\rm dof} = 13.4 / 10$, which statistically indicates a fit of acceptable quality.

The main reason for the fit improvement in the eBW scheme is the reduction in the final proton yields relative to other hadrons, as discussed in the previous section.
Most yields are described on approximately one $\sigma$ level, with the possible exception of $\Lambda$.
In the case of $\Lambda$, the reduction in its yield in the eBW scheme actually leads to somewhat bigger tension with the data, where the yield is measured to be a little higher than the model prediction.
In that regard, we stress that we conservatively include only the established baryonic resonances in our hadron list.
It has been pointed recently that lattice QCD thermodynamics requires the inclusion into HRG of the extra, experimentally uncharted strange baryonic resonances~\cite{Bazavov:2014xya,Alba:2017mqu}, in particular the ones with $|S| = 1$.
Such an inclusion would increase the feeddown contribution to the final $\Lambda$ yields and could reduce the tension with the data.

One should note that the quality of the data for light nuclei is worse than the one for hadrons~(except $\Omega$).
This is seen by the large error bars for the data/model values for light nuclei in Fig.~\ref{fig:fitscentral}.
Therefore, the inclusion of light nuclei into the fit leads to a somewhat artificial decrease of $\chi^2 / {\rm dof}$.
When the yields of light nuclei are excluded from the fit, the extracted $T$ and $V$ values change very little but the reduced $\chi^2$ values are quite different~(see Table~\ref{tab:fits}):
$\chi^2 / {\rm dof} = 15.4/6$ for the zero-width approximation,
$\chi^2 / {\rm dof} = 19.3/6$ for the BW scheme,
and $\chi^2 / {\rm dof} = 9.1/6$ for the eBW scheme.
This elucidates more clearly the significant improvement in the eBW scheme, as the yields of light nuclei, which were removed from the fit, are not affected by modeling of the finite resonance widths.

\subsection{Centrality dependence}

\begin{figure}[t]
  \centering
  \includegraphics[width=.49\textwidth]{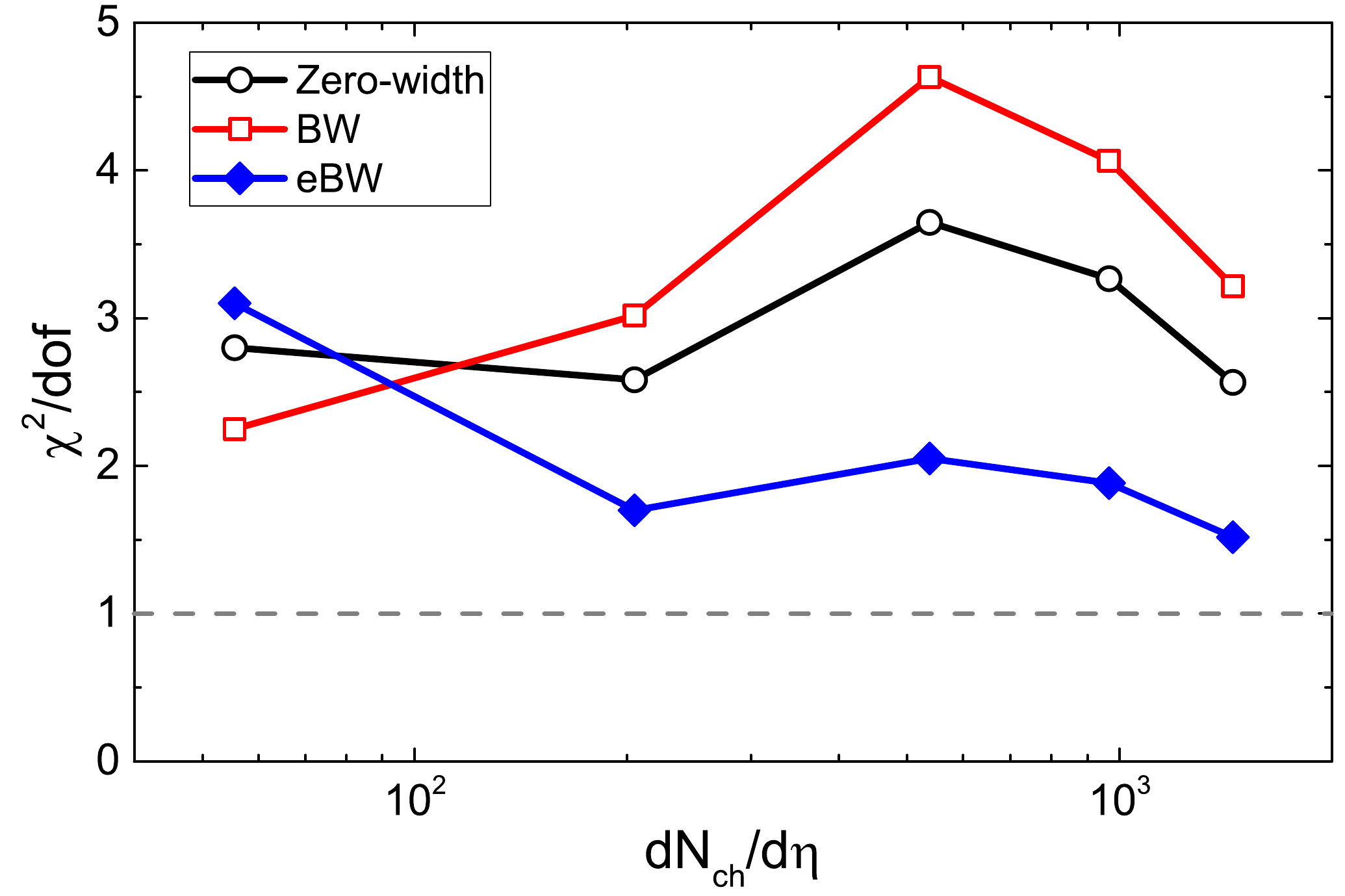}
  \caption{Centrality dependence of $\chi^2/\textrm{dof}$ of thermal fits to Pb-Pb data at $\sqrt{s_{_{NN}}} = 2.76$~TeV performed with the zero-width approximation~(open circles),
  the BW scheme~(open red squares), and the eBW scheme~(solid blue diamonds).
  The lines are drawn to guide the eye.
  }
  \label{fig:chi2cent}
\end{figure}

For the different centralities, 0-10\%, 10-20\%, 20-40\%, 40-60\%, and 60-80\%,
we perform fits to the symmetrized yields of $\pi$, $K$, $K_0^S$, $\phi$, protons, $\Lambda$, $\Xi$, and $\Omega$.
The fit results 
are depicted in Table~\ref{tab:fits}.
The centrality dependence of $\chi^2/{\rm dof}$ is shown in Fig.~\ref{fig:chi2cent}.
The picture at different centralities is generally similar to the one at 0-10\%: the fit quality is notably better in the eBW as compared to the zero-width approximation or the BW scheme.
The only exception is the most peripheral 60-80\% bin, where the best fit quality is obtained in the BW scheme.

\begin{table}
 \caption{Results of the thermal fits to ALICE data for $\sqrt{s_{_{NN}}} = 2.76$~TeV Pb-Pb collisions at different centralities. The fits are performed within HRG model with three different scenarios for resonance widths modeling: the zero-width approximation, the energy independent Breit-Wigner (BW) scheme, and the energy-dependent Breit-Wigner (eBW) scheme.
 The data fitted does not include light nuclei, hence the small difference between the results for the 0-10\% centrality presented in Fig.~\ref{fig:fitscentral} and in this table. } 
 \centering                                                 
 \begin{tabular}{@{\extracolsep{10pt}}cc|ccc}   
 \hline
 \hline
 Centrality & Scheme & \multicolumn{3}{c}{Fit results} \\
 & & $T$~(MeV) & $V$~(fm$^3$) & $\chi^2 / {\rm dof}$  \\
 \hline
 0-10\% & zero-width & $155.3 \pm 2.8$ & $4625 \pm 732$ & 15.4/6 \\
        & BW  & $155.1 \pm 2.8$ & $4506 \pm 718$ & 19.3/6 \\
        & eBW & $157.5 \pm 2.8$ & $4303 \pm 659$ & 9.1/6 \\
\hline
10-20\% & zero-width & $157.1 \pm 2.9$ & $2904 \pm 464$ & 19.6/6 \\
        & BW  & $156.7 \pm 2.9$ & $2866 \pm 459$ & 24.4/6 \\
        & eBW & $159.7 \pm 2.9$ & $2655 \pm 413$ & 11.3/6 \\
\hline
20-40\% & zero-width & $158.7 \pm 2.9$ & $1545 \pm 242$ & 21.9/6 \\
        & BW  & $158.3 \pm 2.9$ & $1517 \pm 239$ & 27.8/6 \\
        & eBW & $161.4 \pm 2.9$ & $1410 \pm 214$ & 12.3/6 \\
\hline
40-60\% & zero-width & $159.3 \pm 2.9$ & $561 \pm 86$ & 15.5/6 \\
        & BW  & $159.0 \pm 2.9$ & $549 \pm 85$ & 18.1/6 \\
        & eBW & $161.4 \pm 2.9$ & $526 \pm 78$ & 10.2/6 \\
\hline
60-80\% & zero-width & $154.4 \pm 2.5$ & $192 \pm 25$ & 16.8/6 \\
        & BW  & $154.7 \pm 2.5$ & $183 \pm 24$ & 13.5/6 \\
        & eBW & $155.2 \pm 2.4$ & $189 \pm 24$ & 18.6/6 \\
\hline
\hline
 \end{tabular}
\label{tab:fits}
\end{table}

Analysis of the centrality dependence of the extracted chemical freeze-out temperature indicates a small increase of the temperature  when going from most central to mid-central collisions, followed by a drop for peripheral collisions.
This behavior appears to be virtually independent of the resonance widths schemes employed in the present work and it is consistent with previous analyses of other authors~\cite{Becattini:2014hla,Chatterjee:2016cog}.

\section{Discussion and conclusions}
\label{sec:summary}

Several remarks are in order in light of the obtained results:

\begin{itemize}

\item Application of the theoretically better motivated energy-dependent Breit-Wigner scheme leads to a notable improvement in the thermal model description of hadron yields at the LHC.
The improvement comes mainly from the reduced feeddown of protons from broad $\Delta$ resonances, due to their suppression at the threshold in the eBW scheme.

\item A smaller but still notable suppression is also seen in the eBW scheme for the yields of $\Lambda$ baryons~(Fig.~\ref{fig:ratios}), which, in contrast to protons, is in tension with the data at the LHC.
It would be useful to clarify the role of the additional strange baryons, not listed in PDG but suggested by lattice QCD~\cite{Bazavov:2014xya,Alba:2017mqu}, which can be expected to yield additional feeddown of $\Lambda$'s.

\item The commonly used way to include the finite widths of the resonances in the thermal model is to apply the energy independent Breit-Wigner scheme.
Our results suggest that more involved modeling is  warranted. In contrast to the eBW scheme, the energy independent BW scheme actually leads to an opposite effect of enhancing the final hadron yields, as seen in Fig.~\ref{fig:ratios}.

\item The so-called 'proton anomaly' at the LHC can be explained, at least partially, within the eBW scheme.
Other explanations, such as
the incomplete hadron spectrum~\cite{Stachel:2013zma,Noronha-Hostler:2014aia},
the chemical non-equilibrium at freeze-out~\cite{Petran:2013lja,{Begun:2013nga,Begun:2014rsa}},
the modification of hadron abundances in the hadronic phase~\cite{Steinheimer:2012rd,{Becattini:2012xb,Becattini:2016xct}},
separate freeze-out for strange and non-strange hadrons~\cite{Bellwied:2013cta,Chatterjee:2016cog},
or the excluded volume interactions~\cite{Alba:2016hwx}, have also been proposed.
We stress that in the present work we stay in the framework of the chemical equilibrium ideal HRG model, the
only difference coming from the modified treatment of resonance widths.
In that regard, our explanation is simpler than the above ones, as it does not introduce any alternative scenarios/interpretations of the data.

\item For some broad resonances, e.g. the $\rho$-meson~\cite{Huovinen:2016xxq}, neither the BW nor the eBW scheme provides an accurate description.
The eBW scheme considered here should not be viewed as an unquestionable improvement over the zero-width approximation or over the BW scheme.
The results should rather be viewed as an illustration of the sensitivity of the thermal model description with respect to the modeling of resonance widths.
In that regard it can also be interesting to consider other approaches, such  as the ones based on the K-matrix formalism~\cite{Wiranata:2013oaa,Dash:2018can} or empirical phase shifts~\cite{Huovinen:2016xxq,Lo:2017lym}.
Analysis within the latter formalism does show indications for a reduced proton yield~\cite{Andronic:2018qqt}, similar to the eBW scheme.

\item Hadronic interactions have been considered here on the level of the resonance formation only.
It is known, however, that thermodynamics of interacting hadrons also receives positive contributions from non-resonant, correlated pairs of interacting constituents.
One prominent example is the P33 channel of the $\pi N$ interaction, where the correlated $\pi N$ pair contribution is notable near the $\Delta(1232)$ production threshold~\cite{Weinhold:1997ig}.
This contribution can negate, to some extent, the proton yield suppression effect obtained in this work for the eBW scheme.
Recent lattice QCD data also favor the existence of the excluded volume type effects in baryon-baryon interactions~\cite{Vovchenko:2016rkn,Vovchenko:2017xad}, which may have a significant effect on thermal model fits~\cite{Vovchenko:2015cbk,Alba:2016hwx,Satarov:2016peb}. It would be interesting to study the above mentioned effects in conjunction with a more elaborate treatment of finite resonance widths advocated here.

\item Relevance of the effects associated with the finite resonance widths modeling can be further tested with an upcoming hadron yield data from the 5.02~TeV Pb-Pb run at the LHC.
The preliminary 5.02~TeV data of the ALICE collaboration~\cite{Bellini:2018khg} has considerably smaller systematic uncertainties compared to the 2.76~TeV dataset, 
and these new data  illustrate a similar overestimation of the proton yields relative to other hadrons within the standard thermal model.
It is predicted here that the eBW scheme will yield a superior fit quality for the finalized 5.02~TeV data over both the zero-width approximation and BW scheme.

\end{itemize}

To summarize, the thermal model description of hadron yields measured in heavy-ion collisions at the LHC is found to be sensitive to the modeling of the finite resonance widths.
The fit quality is improved significantly when the energy dependent Breit-Wigner scheme is applied, the improvement coming mainly from the reduced proton feeddown from the various $\Delta$ resonances.
At the same time, no significant variation in the extracted chemical freeze-out temperature $T$ and volume parameter $V$ is observed.
The results obtained show that a proper modeling of finite resonance widths is important in precision thermal model studies.


\begin{acknowledgments}

V.V. appreciates motivating discussions with the participants of the "Light up 2018 -- An ALICE and theory workshop", June 14-16, CERN, Switzerland.
The work of M.I.G. was supported
by the Program of Fundamental Research of the Department of
Physics and Astronomy of National Academy of Sciences of Ukraine.
H.St. acknowledges the support through the Judah M. Eisenberg Laureatus Chair at Goethe University, and the Walter Greiner Gesellschaft, Frankfurt.

\end{acknowledgments}

\bibliography{bibliography}


\end{document}